\documentclass[aps,prd,superscriptaddress,showpacs,preprint]{revtex4}
\usepackage{graphicx, bm}
\begin{document}
\draft
\title{Probing the $ZZ\gamma$ and $Z\gamma\gamma$ Couplings Through the\\
       Process $e^+e^-\to \nu \bar\nu \gamma$}

\author{ A. Guti\'errez-Rodr\'{\i}guez}
\affiliation{\small Facultad de F\'{\i}sica, Universidad Aut\'onoma de Zacatecas\\
         Apartado Postal C-580, 98060 Zacatecas, Zacatecas M\'exico.\\}
\author{M. A. Hern\'andez-Ru\'{\i}z}
\affiliation{\small Instituto de F\'{\i}sica, Universidad
        Aut\'onoma de San Luis Potos\'{\i}\\
        78000 San Luis Potos\'{\i}, SLP, M\'exico.\\}
\author{M. A. P\'erez}
\affiliation{\small Departamento de F\'{\i}sica, CINVESTAV.\\
             Apartado Postal 14-740, 07000, M\'exico D.F., M\'exico.}

\date{\today}

\begin{abstract}

We study the sensitivity for testing the anomalous triple gauge
couplings $ZZ\gamma$ and $Z\gamma\gamma$ via the process
$e^+e^-\to \nu \bar\nu \gamma$ at high energy linear colliders.
For integrated luminosities of 500 $fb^{-1}$ and center of mass
energies between 0.5 and 1.5 $TeV$, we find that this process can
provide tests of the triple neutral gauge boson couplings of order
$10^{-4}$, one order of magnitude lower than the standard model
prediction.

\end{abstract}

\pacs{14.60.Lm,12.15.Mm, 12.60.-i
\vspace*{2cm}\noindent  E-mail: $^{1}$alexgu@planck.reduaz.mx,
$^{3}$mperez@fis.cinvestav.mx,}

\vspace{5mm}

\maketitle


\section{Introduction}

Triple gauge boson couplings constitute a sensitive probe of
nonstandard interactions \cite{Ellison,Martinez}. In particular,
triple neutral gauge boson couplings (TNGBC) $ZVV$, with $V=Z,
\gamma$, vanish when the three gauge bosons are on mass shell and
Bose symmetry and Lorentz invariance are simultaneously satisfied
\cite{Gaemers,Barroso}. Eventhough these vertices do not reflect
directly the non-Abelian nature of the electroweak gauge group,
since the Yang-Mills sector does not induce them at the tree or
one-loop level, they constitute a window to new physics
\cite{Ellison,Martinez}. These vertices must be induced by loop
effects in any renormalizable theory since they do not possess a
renormalizable structure. At the one-loop level, they can be
generated only by fermionic triangles \cite{Renard,Hernandez}. In
the Standard Model (SM) and the Minimal Supersymmetric Standard
Model (MSSM) they are rather small, of order $10^{-3}$
\cite{Hernandez,Gounaris,Choudhury}, but new heavy quark
generations may enhance them by about one order of magnitude
\cite{Hernandez}.

Experimental measurements of possible TNGBC $ZZ\gamma$ and
$Z\gamma\gamma$ have reached an accuracy at the few percent level
at the Tevatron \cite{Abasov}, even better than the accuracy
obtained for nonstandard deviations of the $WWZ(\gamma)$ couplings
\cite{Abasov1}. Persuasive arguments within the effective
Lagrangian approach indicate that TNGBC are not expected to be
larger than $1\%$ \cite{Ellison,Martinez}. Indirect limits on
these vertices have been obtained from the muon $g-2$ value, the
known bound on the electron electric dipole moment
\cite{Hernandez} and the measurement of the rare decay mode $Z\to
\nu \bar \nu\gamma$ obtained at LEP \cite{Abdallah,Maya}. The
process $e^+e^-\to Z\gamma$ has been studied also at the energies
expected in future linear colliders in order to search for its
sensitivity to the TNGBC through the effects of $Z$ polarization
and initial state radiation \cite{Atag}, as well as possible
CP-violating effects in the decay mode $Z\to \mu^+\mu^-\gamma$
\cite{Perez}.

The process $e^+e^-\to \nu \bar\nu\gamma$ has been studied
previously in connection to the determination of the number of
light neutrino species \cite{Ma} and its possible electric and
magnetic moments \cite{Larios}, as well as in the search of extra
neutral gauge bosons \cite{Aliev}. In the present paper we study
the sensitivity to the TNGBC in the process $e^+e^-\to \nu
\bar\nu\gamma$ at the next generation of linear colliders, namely
the international linear collider (ILC) \cite{Abe} and the compact
linear collider (CLIC) \cite{Accomando}. We expect to get better
limits on the anomalous $ZVV$ couplings through the $Z\to \nu
\bar\nu$ channel because its larger branching ratio induces a
cross section for $e^+e^-\to \nu \bar\nu\gamma$ about a factor 3
larger than the combined rates for $e^+e^-\to Z (\to e^+e^-,
\mu^+\mu^-)\gamma$ \cite{Abadin}. The process $e^+e^- \to Z(\to
\nu \bar\nu)\gamma$ is also competitive with respect to $Z\gamma$
production at the LHC because the inclusive NLO QCD corrections to
the $Z\gamma$ cross section are quite large at high photon
transverse momenta in the SM and reduce the sensitivity to the
TNGBC $ZVV$ \cite{Baur}. We will obtain in the present paper that
the next generation of $e^+e^-$ colliders may reach in fact
sensitivities that are close to the SM predictions for the TNGBC.
In this respect, our  study can be considered complementary to
previous studies on the TNGBC \cite{Atag,Perez}.

This paper is organized as follows: In Sect. II we present the
calculation of the cross section for the process with anomalous
couplings $ZZ\gamma$ and $Z\gamma\gamma$ and, finally, we
summarize our results and conclusions in Sec. III.

\section{CROSS SECTIONS}

The most general anomalous $Z(q_1)\gamma(q_2)Z(P)$ vertex function
is given by \cite{Gaemers}

\begin{eqnarray}
\Gamma^{\alpha\beta\mu}_{Z\gamma Z} (q_1,
q_2,P)&=&\frac{P^2-q^2_1}{M^2_Z}[ h^Z_1(q^\mu_2g^{\alpha\beta}
-q^\alpha_2g^{\mu\beta}) + \frac{h^Z_2}{M^2_Z}P^\alpha(P\cdot q_2
g^{\mu\beta}- q^\mu_2P^\beta )\nonumber\\
&+& h^Z_3\varepsilon^{\mu\alpha\beta\rho}q_{2\rho} +
\frac{h^Z_4}{M^2_Z} P^\alpha\varepsilon^{\mu\beta\rho\sigma}P_\rho
q_{2\sigma} ],
\end{eqnarray}

\noindent where $M_Z$ is the $Z$ boson mass and the respective
$Z\gamma\gamma$ general vertex function can be obtained from Eq.
(1) by the following replacements:

\begin{equation}
\frac{P^2-q^2_1}{M^2_Z} \to \frac{P^2}{M^2_Z}, \hspace{5mm} h^Z_i
\to h^\gamma_i, \hspace{5mm} i=1, 2, 3, 4.
\end{equation}

The form factors $h^V_i$ are dimesionless functions of the $p^2_i$
momenta and in order to avoid violation of partial wave unitarity,
generalized dipole form factors will be considered \cite{Ellison}

\begin{eqnarray}
h^V_i(s)&=&\frac{h^V_{i0}}{(1+s/\Lambda^2)^3}; \hspace{5mm} i=1,
3\\ h^V_i(s)&=&\frac{h^V_{i0}}{(1+s/\Lambda^2)^4}; \hspace{5mm}
i=2, 4.
\end{eqnarray}

We will assume that the new physics scale $\Lambda$ is above the
collision energy $\sqrt{s}$. The Feynman diagrams that contribute
to the process $e^+e^-\to Z(\to \nu \bar\nu)\gamma$ are depicted
in Fig. 1. We will consider only the effect due to the anomalous
$ZV^*\gamma$ vertices for various ILC/CLIC energies and
luminosities. The SM contribution to this process would arise from
initial state radiation coming from the $e^+e^-$ beams and photons
emitted from a virtual $W$ boson exchanged in the $t$-channel. In
order to suppress the SM contribution we impose the cuts $E_\gamma
> 50$ $GeV$ and $45^o <\theta_\gamma <135^o$, where
$\theta_\gamma$ is the polar angle of the photon with respect to
the beam direction \cite{Abdallah,Perez}.

The expression for the respective cross section with anomalous
couplings $ZZ\gamma$ is given by

\begin{eqnarray}
\sigma(e^+e^-\to \nu\bar\nu\gamma)&=&\int\frac{\alpha^2e^2}{256\pi
M^4_Z}C[x_W][(h^Z_{10})^2F_1(s, E_\gamma,
\cos\theta_\gamma)\nonumber\\
&+&(h^Z_{30})^2F_2(s, E_\gamma, \cos\theta_\gamma )] E_\gamma
dE_\gamma d\cos\theta_\gamma,
\end{eqnarray}

\noindent where $E_\gamma$ and $\cos\theta_\gamma$ are the energy
and scattering angle of the photon.

The kinematics is contained in the functions

\begin{eqnarray}
F_1(s, E_\gamma, \cos\theta_\gamma)&\equiv&
\frac{[\frac{1}{2}\sqrt{s}E_\gamma+\frac{1}{2}(s-2)E^2_\gamma -
\sqrt{s}E^3_\gamma - sE^2_\gamma\sin^2\theta_\gamma +
\sqrt{s}E^3_\gamma\sin^2\theta_\gamma] }{(s-M^2_Z)^2+M^2_Z\Gamma^2_Z},\\
F_2(s, E_\gamma, \cos\theta_\gamma)&\equiv& \frac{
[-sE^2_\gamma+2\sqrt{s}E^3_\gamma + \frac{3}{2}
sE^2_\gamma\sin^2\theta_\gamma -
\sqrt{s}E^3_\gamma\sin^2\theta_\gamma]
}{(s-M^2_Z)^2+M^2_Z\Gamma^2_Z},
\end{eqnarray}

\noindent while the coefficient $C$ is given by

\begin{equation}
C[x_W]\equiv \frac{[1-4x_W+8x^2_W] }{x^2_W(1-x_W)^2},
\end{equation}

\noindent where $x_W\equiv \sin^2\theta_W$.

In the case of the anomalous coupling $Z\gamma\gamma$, in order to
get the respective expression for the cross section, we have to
use the substitution given in Eqs. (2-4) and take the appropriated
photon propagator into account. We do not reproduce the analytical
expressions here because they are rather similar to the terms
given in Eqs. (5)-(8).

\section{RESULTS AND CONCLUSIONS}

We expect that the contribution arising from the $ZVV$ anomalous
vertices is enhanced in the process $e^+e^-\to \nu \bar\nu \gamma$
at high energies with respect to the SM contribution
\cite{Perez,Baur}, as it is shown in Fig. 2. In this section we
derive those values of $h^V_{i0}$, $V=Z, \gamma$, which would give
rise to a deviation from the SM prediction at the level of two
standard deviations $(95\%$ C.L.). As it is shown in Eq. (5), the
couplings $h^V_{2, 4}$ do not contribute to the total cross
section \cite{Hernandez} and the calculation of the sensitivity
bounds is facilited by the fact that the CP-conserving and
CP-violating couplings $h^V_{1, 3}$ do not interfere. Even more,
cross sections and sensitivities are nearly identical for equal
values of these couplings \cite{Baur}. For simplicity, we shall
take then one of the vertex contributions at the time. In Figs. 3
and 4 we present the dependence of the sensitivity limits of the
$h^Z_{1, 3}$ vertices with respect to the collider luminosity for
three different values of the c.m. energy. A similar dependence is
obtained for the sensitivity limits of the $h^\gamma_{1, 3}$
vertices with respect to the collider luminosity. In Figs. 5 and 6
we summarize the respective limit contours for these vertices at
the $95\%$ C.L. in the $h^V_{10}-h^V_{30}$ plane for a luminosity
of 500 $fb^{-1}$ and the c.m. energies 0.5-1.5 $TeV$. Finally, in
Tables 1 and 2 we include the $95\%$ C.L. limits obtained for both
vertices for two different values of the new physics scale energy
$\Lambda=3$ and 5 $TeV$ and the collider luminosity of 500
$fb^{-1}$.

\begin{center}
\begin{tabular}{|c|c|c|c|c|}
\hline
                       & \multicolumn{4}{c|}{$\Lambda= 3$ $TeV$}\\
\cline{2-5} $\sqrt{s}$ & $h^Z_{10}$ & $h^Z_{30}$ & $h^\gamma_{10}$ & $h^\gamma_{30}$ \\
\hline
500 $GeV$   & 0.0020   & 0.0016   & 0.0014    & 0.0011  \\
1000 $GeV$  & 0.00098  & 0.0079   & 0.00069   & 0.00056  \\
1500 $GeV$  & 0.00039  & 0.00031  & 0.00028   & 0.00022  \\
\hline
\end{tabular}
\end{center}

\begin{center}
Table 1. Sensitivities achievable at the $95\%$ C.L. for the
$h^V_{1, 3}$ vertices in the process $e^+e^-\to \nu \bar\nu
\gamma$ for different c.m. energies with ${\cal L}=500$ $fb^{-1}$
and $\Lambda= 3$ $TeV$.
\end{center}

\vspace*{5mm}

\begin{center}
\begin{tabular}{|c|c|c|c|c|}
\hline
                       & \multicolumn{4}{c|}{$\Lambda= 5$ $TeV$}\\
\cline{2-5} $\sqrt{s}$ & $h^Z_{10}$ & $h^Z_{30}$ & $h^\gamma_{10}$ & $h^\gamma_{30}$ \\
\hline
500  $GeV$   & 0.0013   & 0.0011   & 0.00097   & 0.00079  \\
1000 $GeV$   & 0.00080  & 0.00065  & 0.00057   & 0.00046  \\
1500 $GeV$   & 0.00037  & 0.00029  & 0.00026   & 0.00021  \\
\hline
\end{tabular}
\end{center}

\begin{center}
Table 2. Sensitivities achievable at the $95\%$ C.L. for the
$h^V_{1, 3}$ vertices in the process $e^+e^-\to \nu \bar\nu
\gamma$ for different c.m. energies with ${\cal L}=500$ $fb^{-1}$
and $\Lambda= 5$ $TeV$.
\end{center}

In conclusion, we have found that the process $e^+e^-\to \nu
\bar\nu \gamma$ at the high energies and luminosities expected at
the ILC/CLIC colliders can be used to probe for nonstandard $ZVV$
vertices. In particular, we can appreciate that the $95\%$ C.L.
sensitivity limits expected for the $h^V_{1, 3}$ vertices at 1-1.5
$TeV$ c.m. energies already can provide tests of these vertices of
order $10^{-4}$, one order of magnitude lower than the SM and MSSM
predictions \cite{Hernandez,Gounaris,Choudhury}. These sensitivity
limits are of the same order of magnitude that those expected in
the process $e^+e^-\to Z (\to l^+l^-)\gamma$, also at future
$e^+e^-$ colliders, through the effects induced by the
polarization of the $Z$ gauge boson and initial state radiation
\cite{Atag}.

\vspace{1.5cm}

\begin{center}
{\bf Acknowledgments}
\end{center}

We acknowledge support from CONACyT and SNI (M\'exico).

\newpage

\begin{figure}[t]
\centerline{\scalebox{0.95}{\includegraphics{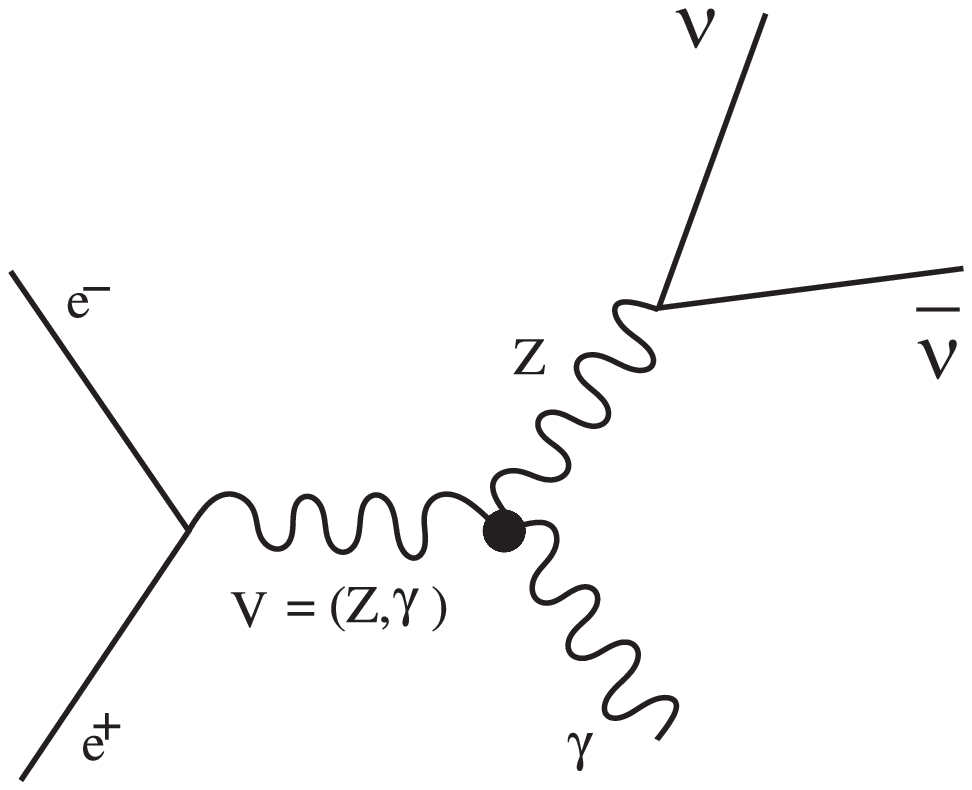}}}
\caption{ \label{fig:gamma} Feynman diagrams induced by the $ZVV$
vertices, $V=Z, \gamma$ in the process $e^+e^-\to \nu \bar\nu
\gamma$.}
\end{figure}

\vspace*{1cm}

\begin{figure}[t]
\centerline{\scalebox{1.15}{\includegraphics{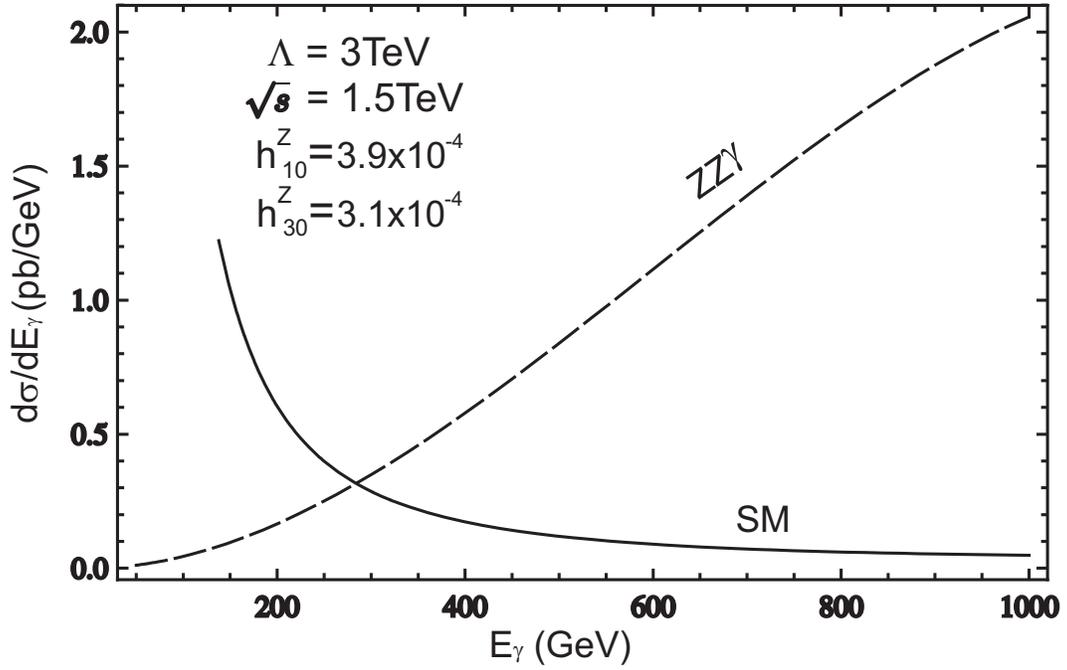}}}
\caption{ \label{fig:gamma} Energy distribution of photon emitted
in the process $e^+e^-\to \nu \bar\nu \gamma$ with $\sqrt{s}=1.5$
$TeV$ and $\Lambda=3$ $TeV$ for the SM and two choices of the
TNGBC $h^Z_{1, 3}$.}
\end{figure}

\begin{figure}[t]
\centerline{\scalebox{1.2}{\includegraphics{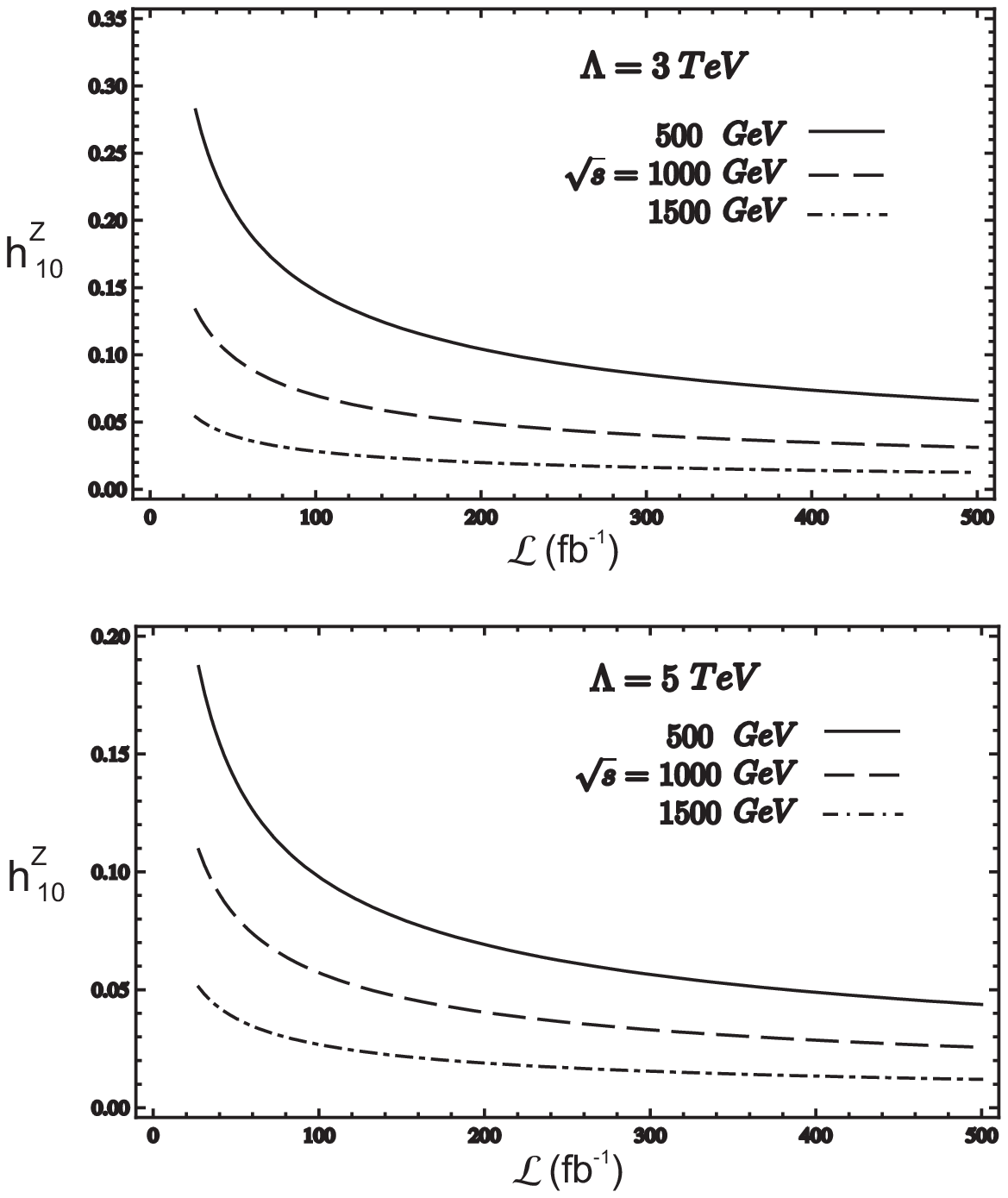}}}
\caption{ \label{fig:gamma} Dependence of the sensitivity limits
at $95\%$ C.L. for the $h^Z_{10}$ vertex for two values of the new
physics energy scale $\Lambda$ and three different values of the
c.m. energy in the process $e^+e^-\to \nu \bar\nu \gamma$.}
\end{figure}

\begin{figure}[t]
\centerline{\scalebox{1.2}{\includegraphics{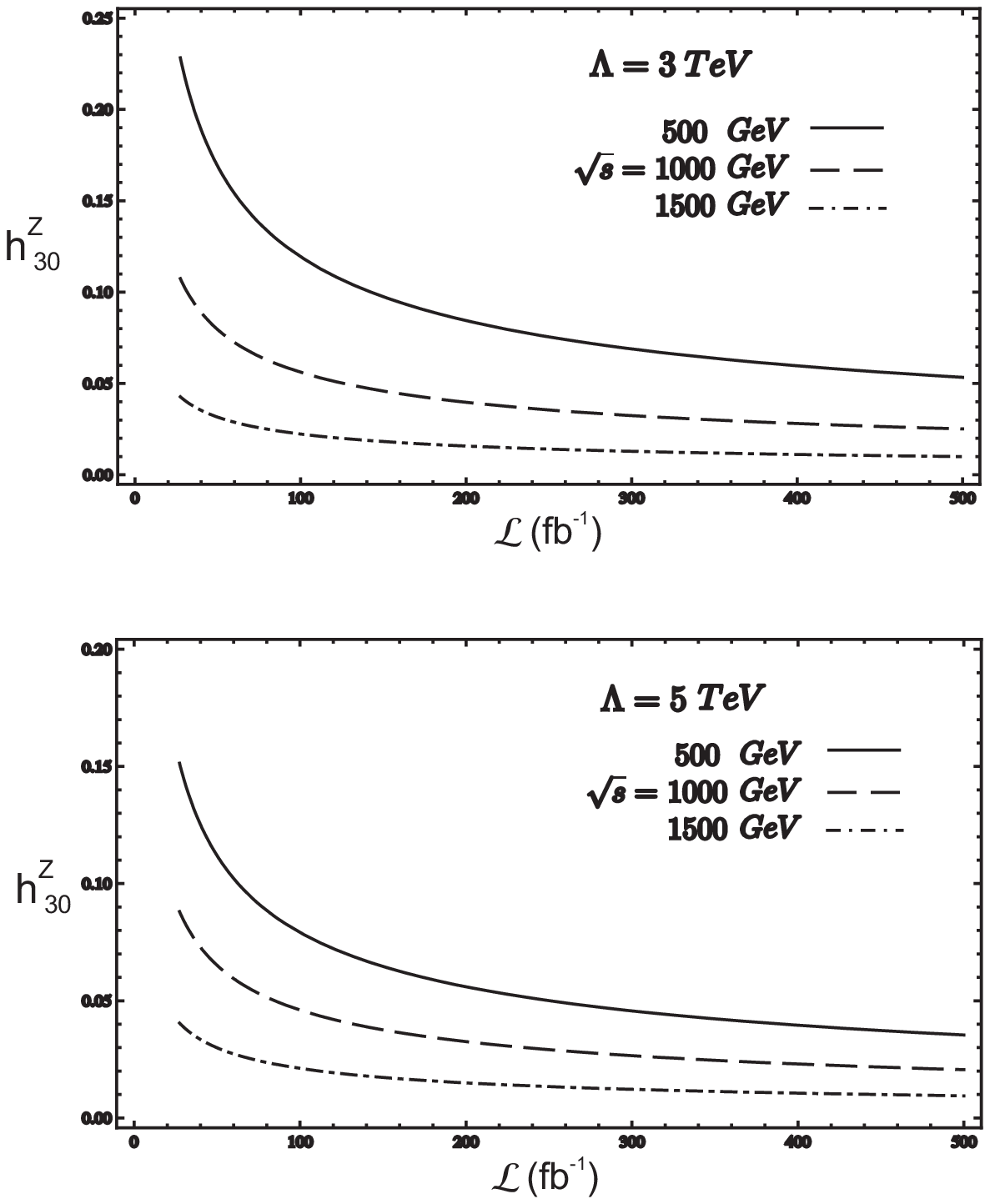}}}
\caption{ \label{fig:gamma} Dependence of the sensitivity limits
at $95\%$ C.L. for the $h^Z_{30}$ vertex for two values of the new
physics energy scale $\Lambda$ and three different values of the
c.m. energy in the process $e^+e^-\to \nu \bar\nu \gamma$.}
\end{figure}

\begin{figure}[t]
\centerline{\scalebox{1.15}{\includegraphics{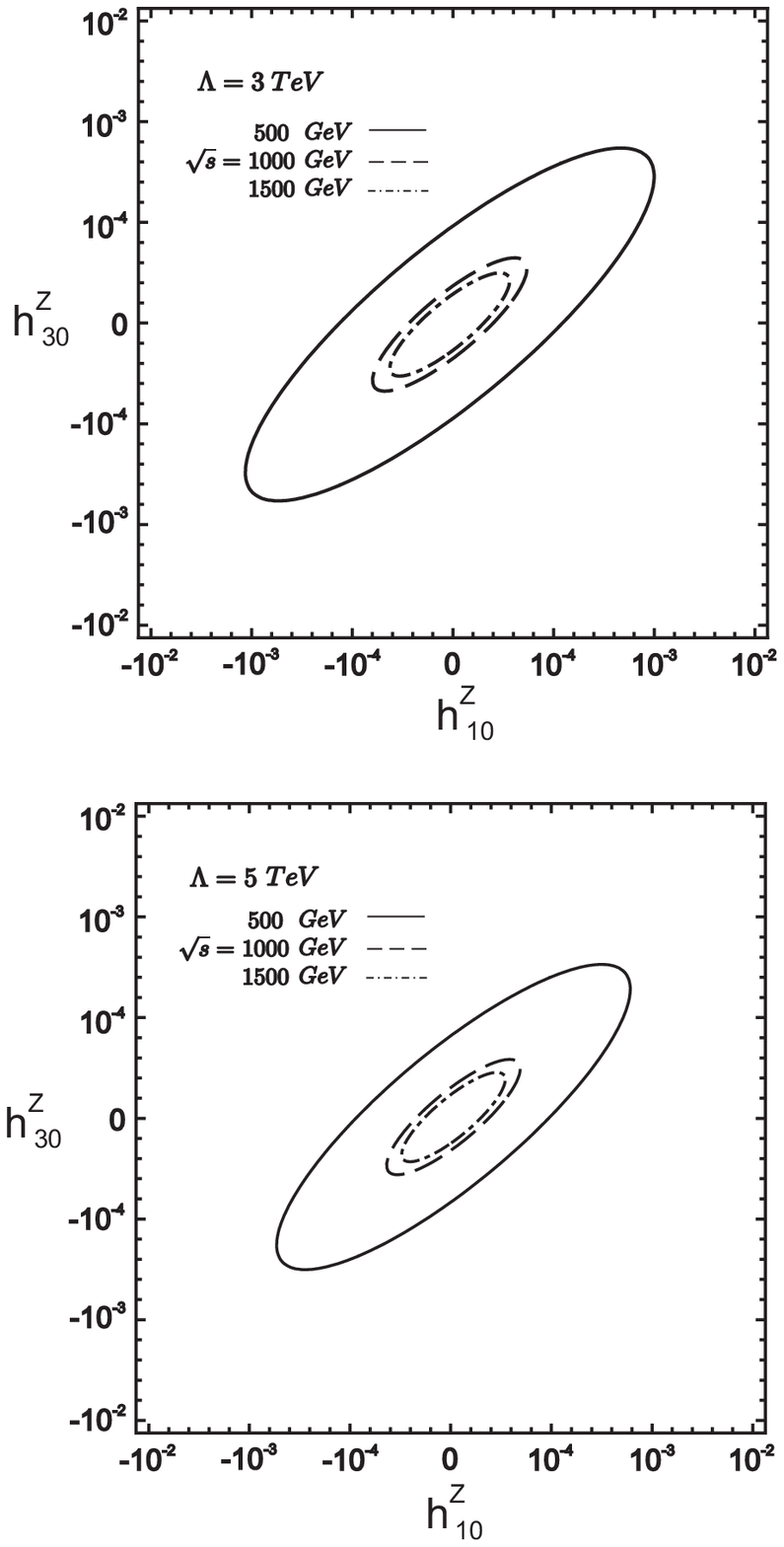}}}
\caption{ \label{fig:gamma} Limits contours at the $95\%$ C.L. in
the $h^Z_{10}-h^Z_{30}$ plane for $e^+e^-\to \nu \bar\nu \gamma$
with a luminosity of 500 $fb^{-1}$ and c.m. energies of 1-1.5
$TeV$.}
\end{figure}

\begin{figure}[t]
\centerline{\scalebox{1.15}{\includegraphics{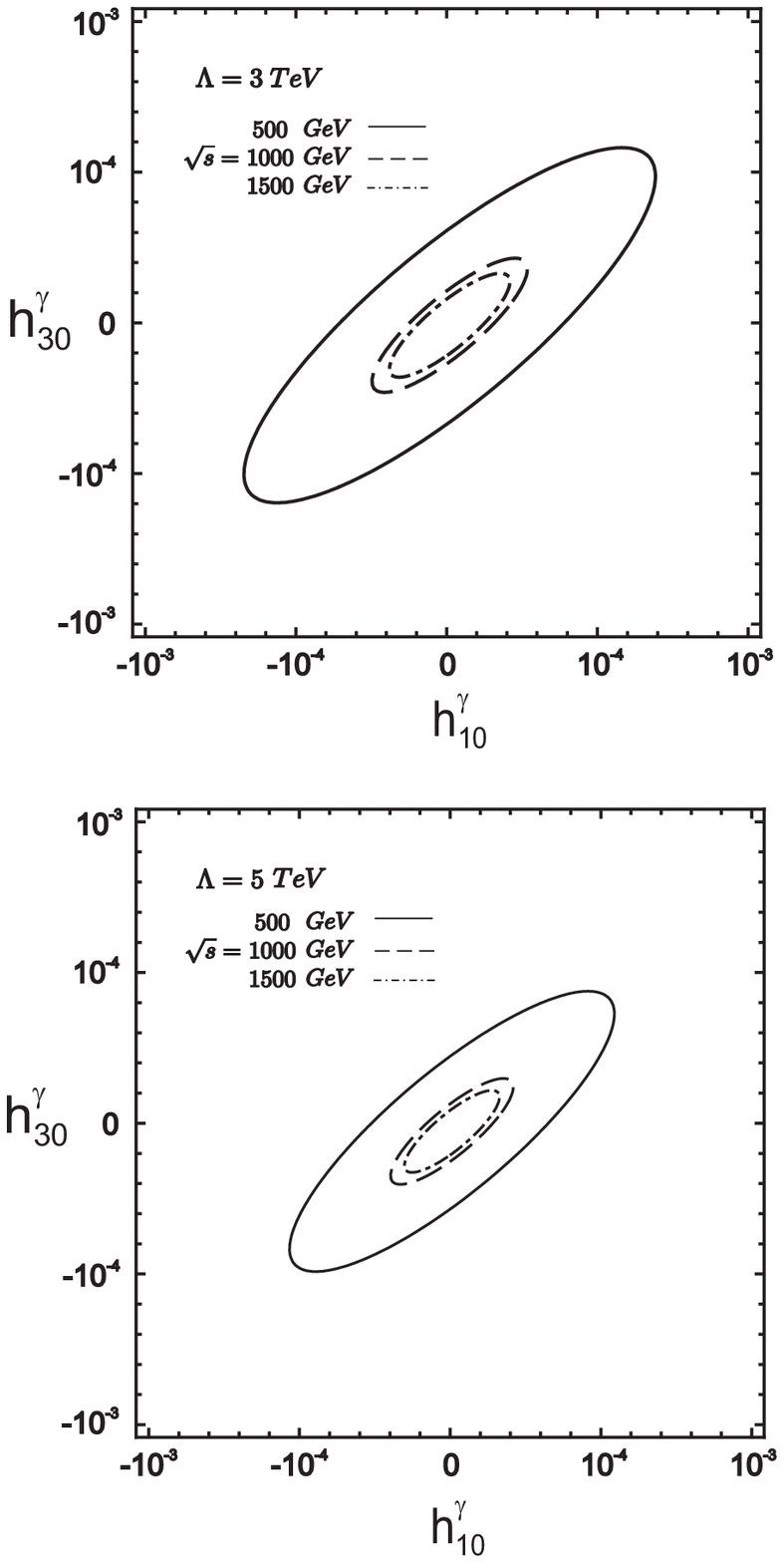}}}
\caption{ \label{fig:gamma} Limits contours at the $95\%$ C.L. in
the $h^\gamma_{10}-h^\gamma_{30}$ plane for $e^+e^-\to \nu \bar\nu
\gamma$ with a luminosity of 500 $fb^{-1}$ and c.m. energies of
1-1.5 $TeV$.}
\end{figure}

\end{document}